\documentclass[twocolumn]{aastex631}

\newcommand\snixf{SN\,2023ixf}
\newcommand\chandra{\emph{Chandra}}
\newcommand\nustar{\emph{NuSTAR}}
\newcommand\swift{\emph{Swift-XRT}}
\newcommand\mdot{$M_\odot\,\rm yr^{-1}$}
\newcommand\msun{$M_\odot$}
\newcommand\kms{km\,s$^{-1}$}
\newcommand\nh{cm$^{-2}$}
\newcommand\lum{erg\,sec$^{-1}$}
\newcommand\cts{counts\,sec$^{-1}$}
\newcommand\flux{erg\,cm$^{-2}$\,s$^{-1}$}

\shorttitle{X-ray evolution of SN 2023ixf}
\shortauthors{Chandra et al.}
\graphicspath{{./}{figures/}}

\begin{document}

\title{
Chandra's
insights into \snixf\footnote{Released on January, 25, 2024}}

\author[0000-0002-0786-7307]{Poonam Chandra}
\affiliation{National Radio Astronomy Observatory,
520 Edgemont Rd, Charlottesville VA 22903, USA}

\author[0000-0002-9117-7244]{Roger A. Chevalier}
\affiliation{Department of Astronomy, University of Virginia,
 Charlottesville VA 22904-4325, USA}

\author[0000-0003-2611-7269]{Keiichi Maeda}]
\affiliation{Department of Astronomy, Kyoto University, Kitashirakawa-Oiwake-cho, Sakyo-ku, Kyoto, 606-8502. Japan}

\author[0000-0002-3356-5855]{Alak K. Ray}
\affiliation{Homi Bhabha Centre for Science Education, TIFR, Mumbai 400088, India }

\author[0000-0002-8070-5400]{Nayana A.J.}
\affiliation{Department of Astronomy, 
University of California, 
Berkeley, CA 94720-3411}

\begin{abstract}

We report  \chandra-ACIS  observations of \snixf\ in M101 on day 13 and 86 since the explosion. The X-rays in both epochs are characterized by   high temperature plasma  from the forward shocked 
region as a result of circumstellar interaction. We are able to constrain the absorption column density at both \chandra\ epochs, which  is much larger than that due to  the Galactic and host absorption 
column, and we attribute it to absorption by the circumstellar matter in the immediate vicinity of \snixf.
Combining our column density measurements with the  published measurement on day 4, we show that the column density declines as $t^{-2}$ between day 4 to day 13 and then evolves as $t^{-1}$. The unabsorbed $0.3-10$ keV luminosity evolves as $t^{-1}$ during the \chandra\ epochs.
On day 13 \chandra\ observation we detect the Fe K-$\alpha$ fluorescent  line at  6.4 keV indicating presence of cold material in the vicinity of the SN. The line is  
absent on day 86, consistent with the decreased column density by a factor of 7 between the two epochs. Our analysis indicates that during 10 years to 1.5 years before explosion, the progenitor was evolving with a constant mass-loss rate of $5.6\times 10^{-4}$\,\mdot. 
%The X-ray measurements indicate asymmetry in the CSM.

\end{abstract}


\section{Introduction} 
\label{sec:intro}

Understanding and mapping the  progenitors to their core-collapse supernovae (CCSNe) is one of the biggest challenges in stellar astronomy.  An important component tying  CCSNe to their massive progenitors is an understanding of stellar mass-loss during the final evolutionary phases, forming the circumstellar material (CSM)  surrounding the star, which can dramatically alter  the observable properties of a supernova (SN). One of the major developments in this field is the indication that most progenitors  undergo the enhanced mass-loss just before explosions, evidence of which has been seen observationally via 
 “flash” spectroscopy \citep[e.g.][etc.]{galyam+14,yaron+17,kochanek19}. While it is not entirely clear, the  mass-loss could be enhanced via several mechanisms, including, but not limited to,  nuclear burning instabilities \citep{sa14} and gravity wave driven mass-loss \citep{qs12}.

\snixf\ is the closest Type II SN observed in decades and, hence, provides an excellent opportunity to study its properties in detail. It was discovered  on 2023 May 19.73 UT in the nearby galaxy M101 by an amateur astronomer Koichi Itagaki  \citep{itagaki+23} and was quickly classified as a Type II supernova \citep{perley+23}. Later it was further classified as a Type IIL SN \citep{bianciardi+23}.
\citet{hiramatsu+23} estimated the explosion date to be  MJD $= 60082.743 \pm 0.083$ (2023 May 18.75 UT), corresponding to 0.98 days before the discovery. We adopt this explosion date throughout the paper. 

 Due to the proximity, detailed observations of \snixf\ have been possible in several wavebands.  In addition, M101 is a well observed galaxy with a rich archival pre-explosion dataset. The pre- and post-explosion dataset have provided important insights into the progenitor of \snixf. The pre-explosion images have revealed the progenitor star
to be a red supergiant in a dusty region \citep{jencson+23,kilpatrick+23,soraisam+23,vandyk+23}. Several groups have constrained the mass of the progenitor, which  have resulted so far in a large mass range within 8--20\,\msun\
\citep{pledger+23,kilpatrick+23,jencson+23,soraisam+23,niu+23}.

The optical light curve of  \snixf\ is characterized by a sharp rise to peak in 5 days from absolute magnitude $-M_V=10$ mag to 
$M_V=-18$ and then a plateau phase lasting for 30 days at magnitude $-17.6$  followed by a smooth decay  \citep{hiramatsu+23,teja+23}.
The data have revealed prominent  flash features of hydrogen, helium, carbon, and nitrogen in the spectra up to 5 days  \citep{yamanaka+23, jg+23, smith+23, bostroem+23,teja+23, hiramatsu+23}. The UV data along with the flash features indicate a temperature rise which is not expected in a pure-shock cooling event, indicating a delayed shock breakout in a dense CSM \citep{zimmerman+23, hiramatsu+23}. 
%This happens when the CSM is so dense that the shock breakout happens in the CSM instead of photosphere. 
\citet{vasylyev+23} measured high polarization and rapid evolution during 2-15 days, 
%with 1\%, decreasing to 0.5\% day 3.5 onwards, combined with rapid change in the polarization angle during $+2.5$ to $+4.5$. This coincides with disappearance of flash features, and
%can be 
and  explained  it as asymmetric ejecta coming out of dense asymmetric CSM, suggesting a highly asymmetric mass-loss process.

 The early X-ray emission was reported by \nustar\ within 4 days  \citep{grefenstette+23}, whereas the first radio detection occurred on 2023 Jun 17, around a month after explosion \citep[10 GHz flux density $41 \pm 8$ $\mu$Jy,][]{vla+23}. \citet{berger+23} reported a non-detection in  millimeter bands at  2.6--18.6 days post-explosion.

There are various mass-loss estimates in the literature. Based on flash ionization features,  \citet{jg+23,zimmerman+23, bostroem+23} have determined an enhanced  mass-loss rate of $\sim 10^{-2}$\,\mdot\ just before explosion and indicated the existence of the confined CSM at
$<10^{15}$\,cm.  The light curve modeling has provided even a higher mass-loss rate \citep[e.g.,][]{hiramatsu+23}. Based on H-alpha evolution, \citet{zhang+23} have estimated mass-loss rate of $6\times 10^{-4}$\,\mdot, 2--3 years before explosion. 
\citet{soraisam+23} found a mass-loss rate of $(2-4)\times 10^{-4}$\,\mdot\, using  19 years of archival IR data, which was also obtained by \citet{grefenstette+23} in their X-ray \nustar\ measurements.
 On the lower end, \citet{jencson+23} and \citet{niu+23} estimated
 it to be $(3-30)\times10^{-5}$\,\mdot, and $10^{-5}$\,\mdot, respectively, in $3-20$ years before explosion. 
The metallicity has ranged between half the solar
\citet{niu+23} to solar \citep{vandyk+23,zimmerman+23}. 

We present here \chandra\ X-ray observations of \snixf\ providing us the opportunity to constrain the evolution of the progenitor.
 We adopt a distance of 6.85 Mpc \citep{riess+22}, a Milky Way extinction of $E(B - V )_{\rm MW}  =
 0.0077$ mag \citep{sf11}, and a host galaxy extinction of $E(B - V )_{\rm host}  = 0.033$ \citep{lundquist+23, smith+23,jg+23, hiramatsu+23}. We also adopt  a standard extinction law with $R_V = 3.1$ \citep{cardelli+89}.

\section{Observations}

\snixf\ was observed with the  \chandra\ under the approved DDT program 24508911 (PI: Chandra).
The first observations  took place 
on 2023 May 31 15:34:04 UT for 20.3 ks. The second set of observations was conducted on 2023 Aug 11 18:17:07 and 2023 Aug 12 12:07:33 UT with 11.18 and 10.02 ks, respectively.
The observations were taken with ACIS-S without grating in a VFAINT mode.  
We extracted the spectra using CIAO 
software\footnote{\url{http://asc.harvard.edu/ciao/}} version 4.15 and
used {\tt xspec} under 
HEAsoft\footnote{\url{http://heasarc.gsfc.nasa.gov/docs/software/lheasoft/}} version 6.31
to carry out the spectral analysis. 
For the second epoch, the individually processed spectra on Aug 11 and 12 were combined using  {\tt combine\_spectra}. A total of 720 counts were obtained
in May 2023 spectrum and 233 counts in Aug 2023 in about 20 ks of exposure at each epoch.

\begin{figure}
\centering
\includegraphics*[angle=270, width=0.46\textwidth]{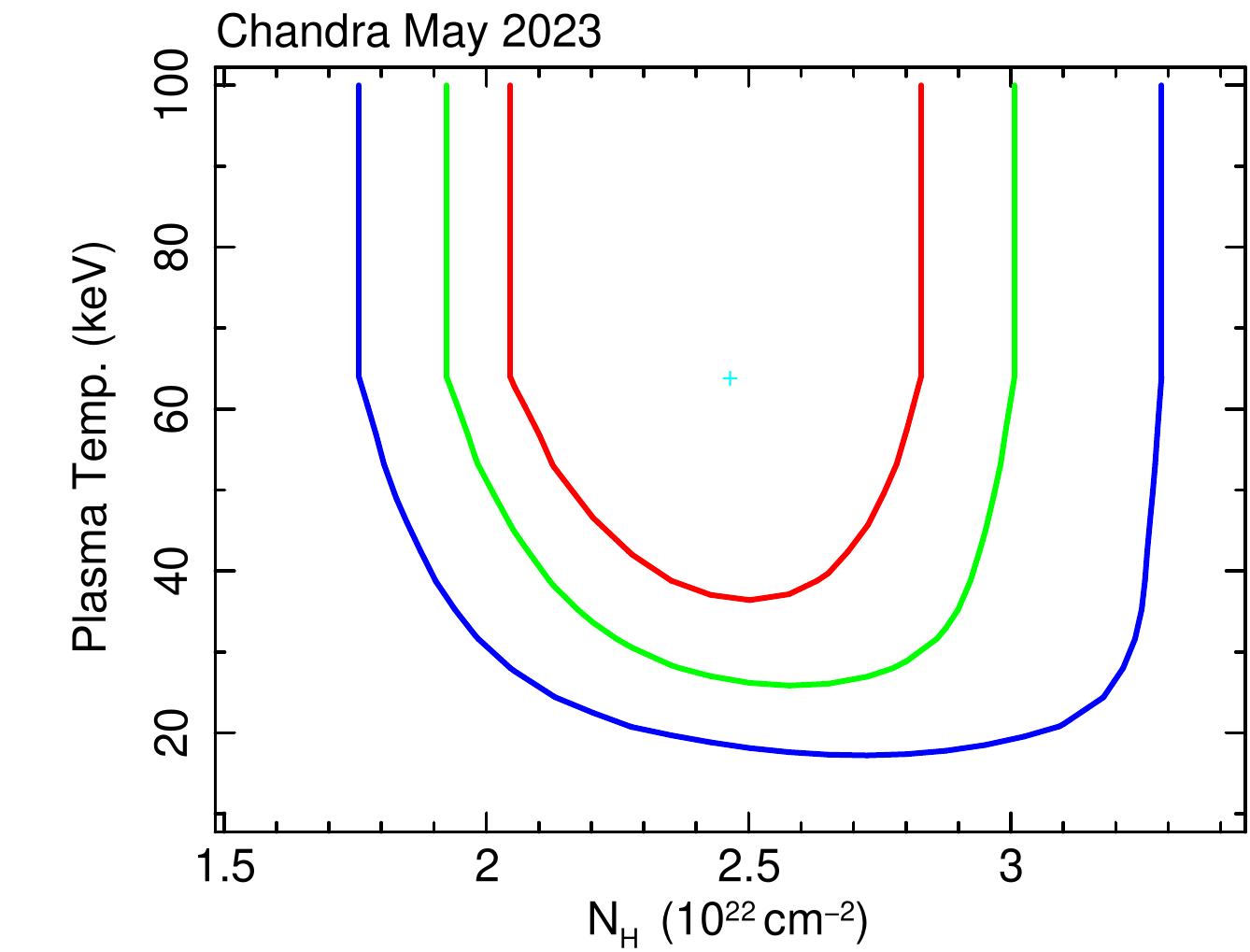}
\includegraphics*[angle=270, width=0.46\textwidth]{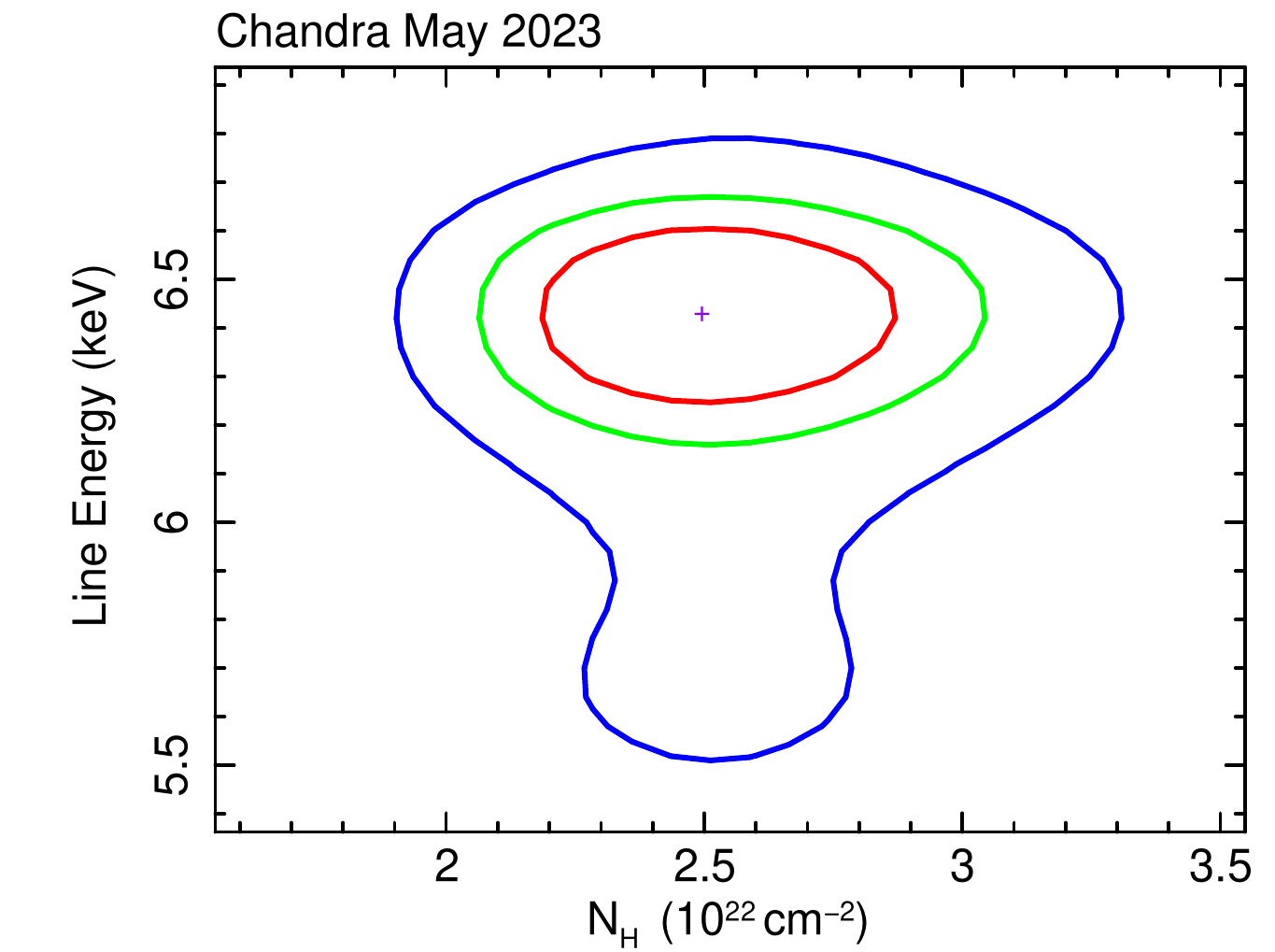}
\caption{{\it Left:} Contour plot of the best fit parameter values of the  column density and the plasma temperature for the May 2023 \chandra\ spectrum. The column density is well constrained while the temperature is not. {\it Right:}   Contour plot of the best fit values of the Gaussian line energy and the column density. Both  parameters are well constrained.
\label{fig:may}}
\end{figure}

For the May 2023 spectral fits, initially we  fit the optically thin ionized plasma model absorbed with a column density. 
We fit the data assuming solar metallicity.
While the column density is well constrained, the best fit temperature ($\sim 60$ keV)  is not constrained towards the upper end (Fig. \ref{fig:may}). That is most likely the case when the plasma temperature is high as the  \chandra\  energy range is not sensitive to such high plasma temperature. We note that \citet{grefenstette+23} constrained the plasma temperature to 34\,keV at around 29th May based on \nustar\ data (day 11), which is within 2 days of our first epoch. Thus we fix  the plasma    temperature to 34 keV in our fits.
 We also see an excess of flux around $6-7$\,keV, though not captured in the plasma model. We add a Gaussian to fit the excess flux. While the line width is not constrained 
due to poor energy resolution of \chandra\, the best fit line energy is $\sim 6.4$\,keV, which matches with the Iron K$\alpha$ line energy, and is also seen in the \nustar\ data. 
We  freeze the line width to 
0.2\, keV
(see \S  \ref{sec:results} for the details). 
The line energy is $6.43^{+0.19}_{-0.20}$\,keV (Fig. \ref{fig:may}).  Our best fit column density is $2.50^{+0.40}_{-0.34} \times 10^{22}$ \nh. Our value is slightly smaller than that in the \nustar\ data 
on day 11 $(5.6\pm2.7)\times10^{22}$\,\nh), though consistent within error bars.

\begin{figure*}
\centering
\includegraphics*[angle=270, width=0.49\textwidth]{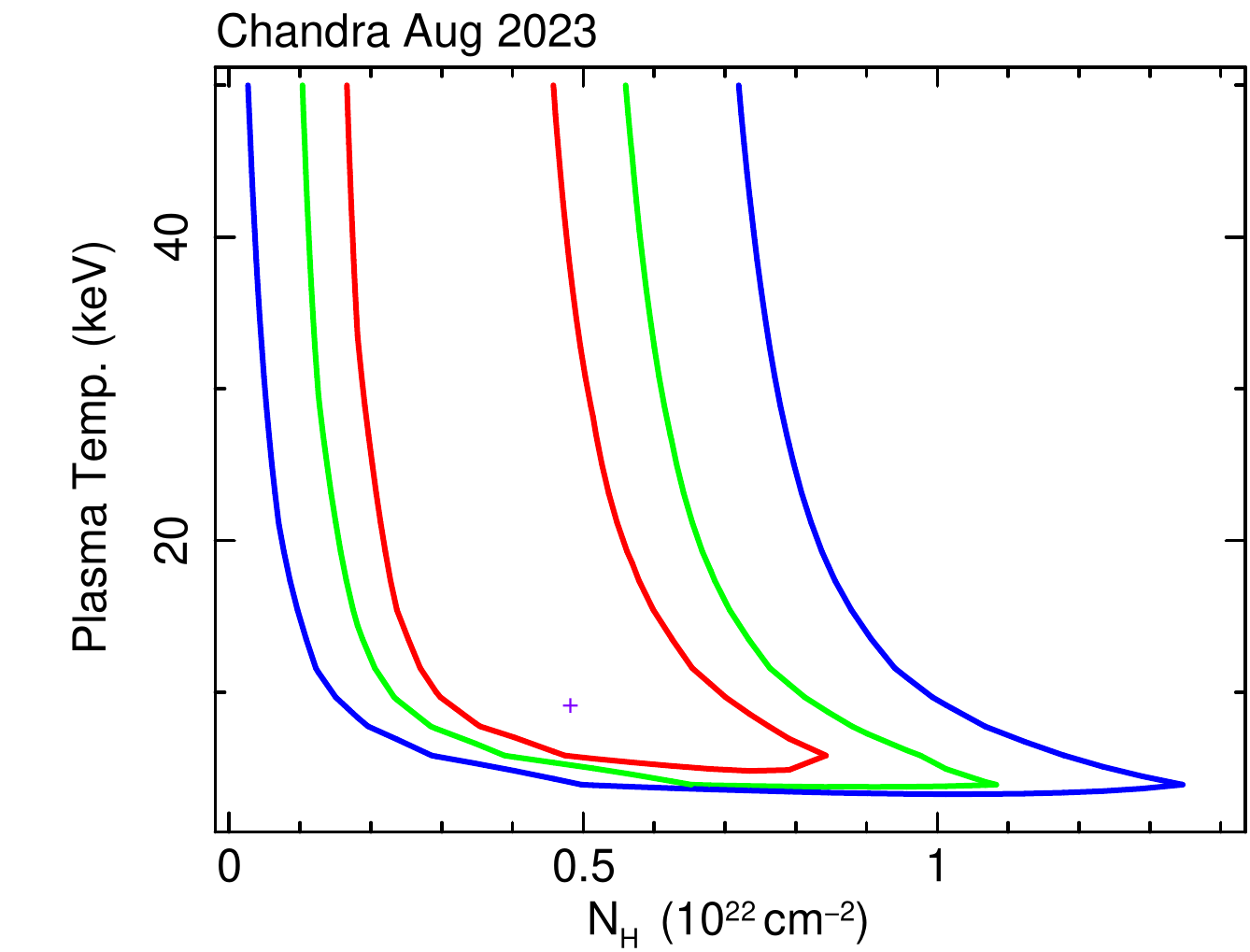}
%\vspace*{-1cm}
\caption{Contour plot of the best fit parameter values for the column density and the plasma temperature for the Aug 2023 \chandra\ spectrum. The column density is well constrained while the temperature is not.  
\label{fig:aug}}
\end{figure*}

\begin{figure*}
\centering
\includegraphics*[angle=270, width=0.49\textwidth]{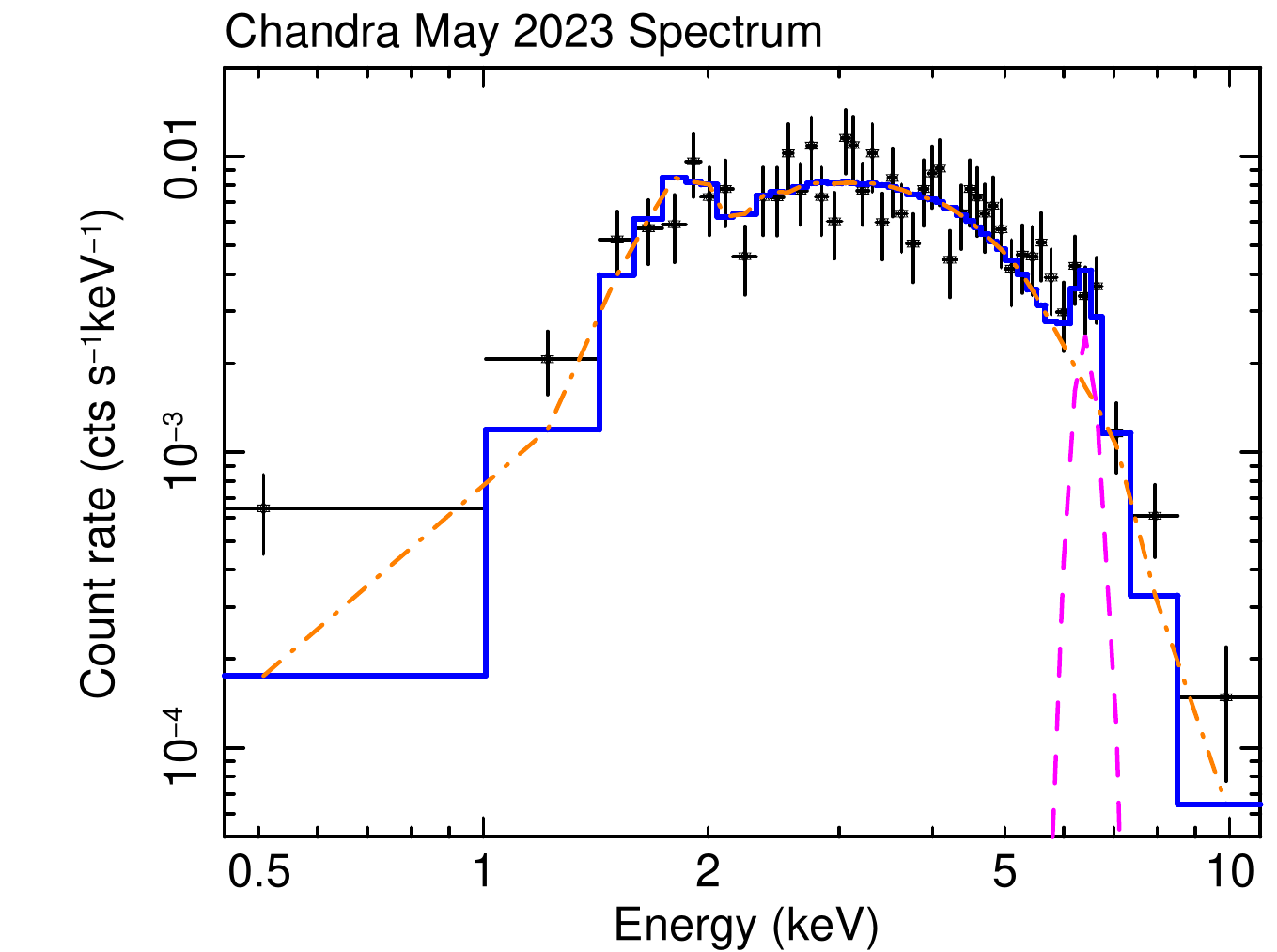}
%\vspace*{-1cm}
\includegraphics*[angle=270, width=0.49\textwidth]{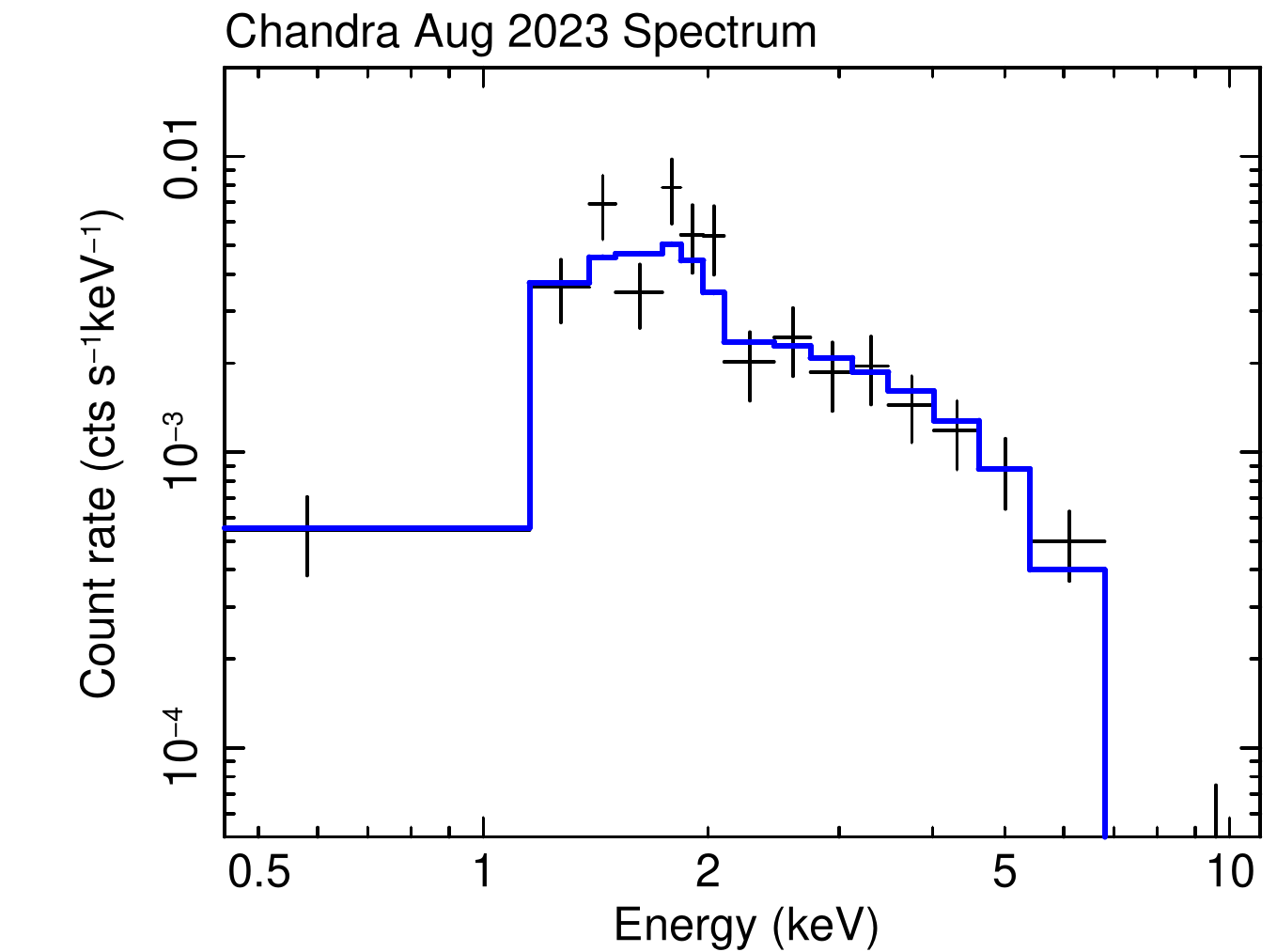}
\caption{Left: May 2023 best fit spectrum of \snixf. The model is fit with an absorbed plasma with a Gaussian profile. Right: Aug 2023 best fit spectrum of \snixf.
The 6.4 Fe K$\alpha$ line is no longer detected.  
\label{fig:spec}}
\end{figure*}

In the second spectrum, we again start with fitting the absorbed plasma model and allow both the temperature and column density to vary.  
There may be a hint of temperature becoming lower than the first epoch  (best fit temperature $\sim 10$ keV); however, due to limited energy range of \chandra, it is not possible to constrain the upper bound of the temperature (Fig. \ref{fig:aug}).
As discussed in 
 \S \ref{sec:results},  we freeze the temperature to 21 keV.  The column density is  well constrained  at this epoch and has substantially decreased to $0.36^{+0.22}_{-0.17} \times 10^{22}$ \nh (Fig. \ref{fig:spec}). We do not see  evidence of the 6.4\,keV line at this epoch. We discuss this further in \S \ref{sec:results}.
%$1.9\times
%$10^{-14}$\, erg\,cm$^{-2}$\,s$^{-1}$\,keV$^{-1}$, consistent with a non-detection.}
In Fig. \ref{fig:spec}, we plot the best fit spectra at both epochs.

\section{Results and discussion}
\label{sec:results}

\subsection{SN 2023ixf X-ray characteristics}

Table \ref{tab:xspec} lists the best fit model parameters and the derived flux values.
Due to the limited energy range, the \chandra\ observations could not constrain the plasma temperature. 
We use  the temperature  of 34 keV in the first spectrum, as obtained in   the nearly coincident   \nustar\ data \citep{grefenstette+23}. 
As discussed below, we  find that during the \chandra\ observation epochs, the column density and the luminosity evolution  follow the standard wind scenario under constant mass-loss rate
 $\rho\propto r^{-2}$ \citep{cf17}. Based on this we scale temperature to $t^{-2/(n-2)}$  \citep[where $n$ is the power-law index of the ejecta density profile][]{cf17}. 
 
 If we assume $n
\sim 10$, the temperature should evolve as $34$ keV $(t/13 \ {\rm day})^{-0.25}$. Thus at the second epoch the temperature should  be 21 keV.  While the temperature may only be approximate, this does not introduce significant errors in our modeling.
We find that for a 20\% change in the plasma temperature, the column density changes by only 5\% and the flux normalization changes by 1.5\%

The high plasma temperature  likely means that the X-ray emission is arising from the forward shock. We do not see a low energy plasma. It is possible if the CSM density is high enough that a cool dense shell forms between the
forward and reverse shocks and most of the lower temperature emission from the reverse shock is  absorbed. 
 In the forward shock interpretation, the shock velocity can be deduced as $v_{\rm sh} = (16 kT/ (3 \mu))^{1/2} = 5000(kT /34\,{\rm keV})^{1/2}$\,\kms, where $k$ is Boltzmann’s constant and $\mu$ is the mean particle weight which is taken to be 0.61 here. 
 Our velocity estimate is slightly smaller than the value of 8500 \kms\ derived by \citet{jg+23}, based on H$\alpha$ and H$\beta$ line profiles. This suggests that either the H$\alpha$ and H$\beta$ profiles are arising from physically distinct regions than the X-ray emission or the shock velocity is evolving to a lower value, since \citet{jg+23} derived the above ejecta velocity at earlier epochs.

\begin{deluxetable}{lcc}
\tabletypesize{\footnotesize}
\tablecaption{Spectral model fits to the \snixf\ spectra
\label{tab:xspec}}
\tablewidth{0pt}
\tablehead{\colhead{} & \colhead{Spectrum 1} &  \colhead{Spectrum 2}}
\startdata
Date & 2023 May 31&  2023 Aug 11--12\\
Epoch & day $\sim 13$ & day $\sim 86$\\
Exposure (ks) & 20.3 & 21.2\\
Model & $N_{\rm H,tot}\times$(apec+Gauss) & $N_{\rm H,tot} \times$ apec\\
$N_{\rm H, tot}$\,(\nh) & $2.50^{+0.40}_{-0.34}\times10^{22}$ & $0.36^{+0.22}_{-0.17} \times 10^{22}$ \\
apec kT (keV)  & 34 (fixed)& 21 (fixed)\\
apec flux (\flux)& $1.43^{+0.09}_{-0.09} \times 10^{-12}$  & $2.09^{+0.23}_{-0.23} \times 10^{-13}$ \\
Gauss kT (keV)  & $6.43^{+0.19}_{-0.20}$ & $\cdots$\\
Gauss Sigma (keV)  & 0.20 (fixed)& $\cdots$\\
Gauss flux  (\flux) & $1.01^{+0.45}_{-0.45} \times 10^{-13}$ & $\cdots$ \\
Abs. flux (\flux) & $1.01^{+0.07}_{-0.07} \times 10^{-12}$ & $2.09^{+0.23}_{-0.23} \times 10^{-13}$ \\
Unabs. flux (\flux)& $1.53^{+0.10}_{-0.10} \times 10^{-12}$ &  $2.55^{+0.28}_{-0.28} \times 10^{-13}$\\
Unabs. $L_{\rm 0.3-10\,keV}$ (\lum) & $8.53^{+0.56}_{-0.56} \times 10^{39}$ &  $1.43^{+0.16}_{-0.16} \times 10^{39}$\\
${\chi}^2/{\nu}$ & 1.04 & 0.65 \\
\enddata
\end{deluxetable}

\begin{figure*}
\centering
\includegraphics*[angle=0, width=0.49\textwidth]{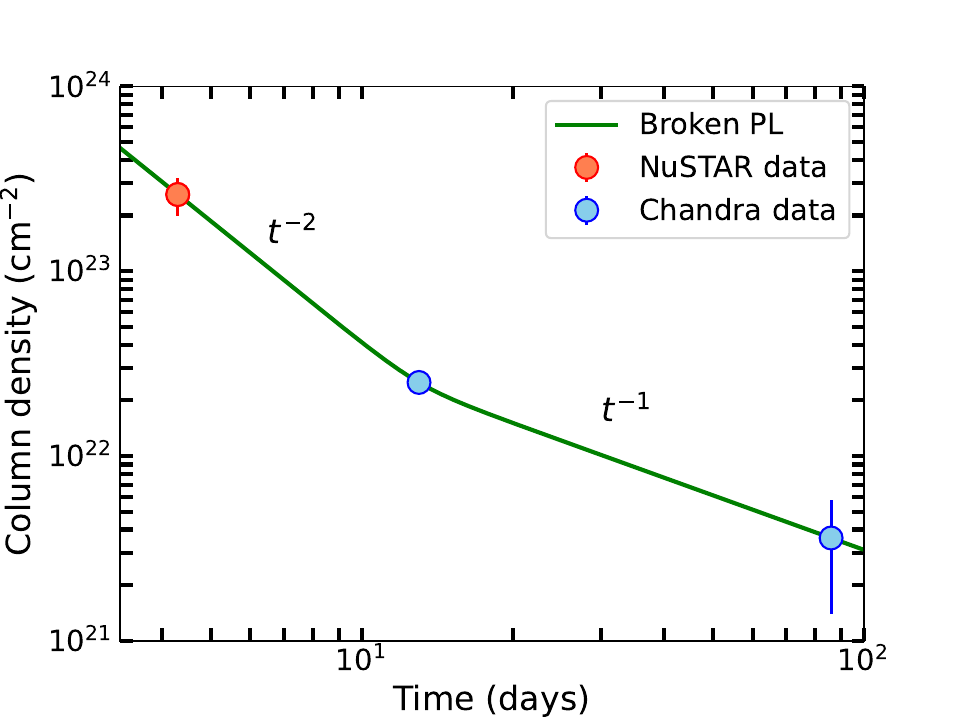}
\includegraphics*[angle=0, width=0.49\textwidth]{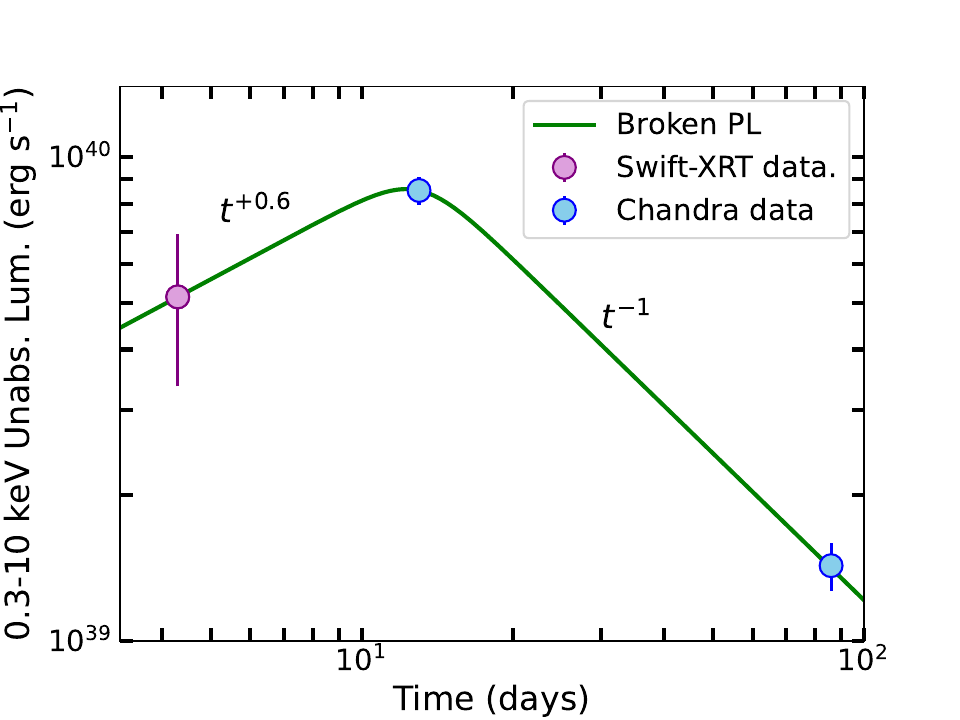}
\caption{Evolution of column density (left) and 0.3--10 keV unabsorbed luminosity (right). The first red data point on day 4 in the  column density  plot is  the \nustar\ data point \citep{grefenstette+23}. The purple data point on the luminosity evolution plot is the \swift\ data point. While the column density evolves faster between day 4 and 13 before starting to follow the standard $t^{-1}$ evolution, the 0.3--10\, keV luminosity initially rises and then follows $t^{-1}$ evolution.
\label{fig:nh}}
\end{figure*}

The extinction due to the Milky Way and the host is $E(B - V )_{\rm MW}  =
 0.0077$ and $E(B - V )_{\rm host}  = 0.033$, respectively. We adopt the method listed in \citet{schlegel+98}, recalibrated by \citet{sf11}, and derive
 $N_{\rm MW}  =0.45\times10^{20}$\,\nh\ and $N_{\rm host}  = 1.91\times10^{20}$\,\nh.  Thus the total column density due to Milky Way and M101 is $2.36\times10^{20}$\,\nh. The column densities derived at the two \chandra\ epochs are $2.50 \times 10^{22}$ \nh\ and $0.36 \times 10^{22}$ \nh, respectively, which are around 130 and 20 times larger than that in the direction of M101. Thus 
 most of the column density is due to the presence of 
 not-completely-ionized material in the vicinity of the CSM. 
  The column density evolves as $t^{-1.03\pm0.29}$ between days 13 and 86, suggesting that the wind corresponding to the CSM seen between these epochs evolved with the standard $r^{-2}$ dependence, where $r$ is the CSM distance from the SN explosion center. However, if one combines this along with the column density obtained in the \nustar\ early data on day 4, it appears that the column density decreased faster between day 4 and 13 with $t^{-2}$ before starting to follow $t^{-1}$
evolution (Fig. \ref{fig:nh}). 
This is consistent with the confined dense CSM seen via the flash ionization spectroscopy \citep{jg+23, bostroem+23, teja+23, smith+23, hiramatsu+23}, 
and is indicative of a scenario in which  the supernova underwent a very high mass-loss rate just before the explosion which resulted in a high column density on day 4, before  transitioning into a low density region of a nearly constant mass-loss rate revealed around day 13 onwards (see below). 
The column density on 
day 4 from the \nustar\ data 
corresponds to an electron scattering column depth of $\sim 0.2$.
Thus electron scattering may have been important before day 4, but not after day 4.

The flux between the two \chandra\ epochs changes by a factor of $\sim 6$.  The unabsorbed 0.3--10.0 keV luminosity evolves at $t^{-0.95\pm0.07}$ which is  also consistent with the standard 
evolution \citep{cf17}. The luminosity on day 13 is $8.53\times10^{39}$\,\lum\, which is slightly lower than the extrapolated value in the \chandra\ energy range
from the  \nustar\ observation at day 11 ($\sim1.4 \times 10^{40}$ \lum), but is roughly consistent within the error bars. 
\citet{panjkov+23} reported the first \swift\
detection on day 4.3 and combined the first 46 days of data to derive  an unabsorbed luminosity of $ (3.0\pm0.7)\times 10^{39}$\,\lum. 
Since we know that parameters such as column density are evolving quickly, we do not take  this approach. 
We combine the \swift\ measurements covering \nustar\ epoch 1, i.e. day 3.9--4.8 and find that there is a 3-$\sigma$ detection with 0.3--10\,keV count rate of
$(1.88 \pm 0.65)\times 10^{-3}$\,\cts. 
 We derive the flux using the \nustar\ best fit parameters on day 4, i.e. $N_H=2.6\times 10^{23}$\,\nh\ 
and  use a plasma temperature of 34 keV \citep{grefenstette+23}.
This provides an unabsorbed (absorbed) 0.3--10 keV 
flux of 
$(9.16 \pm 3.17) \times 10^{-13}$ \flux\ ($(3.07 \pm 1.06) \times 10^{-13}$ \flux), or an
equivalent unabsorbed 0.3--10 keV luminosity of $(5.14\pm1.78)\times 10^{39} $\,\lum.  As we show in Fig. \ref{fig:nh}, the 0.3--10\,keV unabsorbed luminosity initially increases between day 4 and 13 as $t^{+0.6}$ and then evolves as $t^{-1}$ throughout. This simply means that initially the plasma is quite hot and more emission is coming in the hard X-rays. The X-ray behaviour is qualitatively consistent with the one reported in \citet{zimmerman+23}.

Assuming that the radiation is adiabatic at \chandra\ epochs on days 13 and 86, we use equation 3.8 of  \citet{fransson+96} to calculate the mass-loss rate. We estimate the spectral luminosity at 10\,keV and derive the Gaunt factor at 10 keV using \citet{mewe+86}. The mass-loss rates are 
$(5.58\pm 0.18) \times 10^{-4}$\,\mdot and 
$(5.63\pm 0.30) \times 10^{-4}$\,\mdot, on days 13 and 86, respectively. 
The uncertainties are only due to the flux errors. 
Here we assumed the forward shock velocity of $\sim5000$\,\kms\  based on the X-ray temperature, and the wind velocity of 115\,\kms\ based  on the narrow optical line widths \citep{smith+23}. 
The constant mass-loss rate between the two \chandra\ epochs is consistent with the scenario where the forward shock is moving in a wind profile of $\rho_{\rm CSM} \propto r^{-2}$.

We check whether our assumption of forward shock being adiabatic is correct. 
In the radiative case, the luminosity of the  forward shock will be $\propto \dot M$, whereas the adiabatic shock luminosity will scale as $\propto \dot M^2$. Using \citet{cf17}, we can estimate the cooling time for the forward shock.
The cooling time scale
as compared to SN age (assuming $n=10$ and forward shock velocity $5000$\,\kms) is $t_{\rm cool}/t_{\rm age}=0.44 t_{\rm age, d}/(\dot M_{-4}/v_{\rm wind,2})$. Here $t_{\rm age, d}$ is the SN age in days, $\dot M_{-4}$
is the mass-loss rate in units of $10^{-4}$ \mdot, and $v_{\rm wind,2}$ is the wind velocity in units 100 \kms. This means the forward shock is radiative on day 4, if the mass-loss rate is larger than $2.2 \times10^{-4}$\,\mdot.  From the flash ionization spectral analysis at this epoch \citep[e.g.][]{jg+23, bostroem+23, hiramatsu+23} and the delayed shock breakout calculation \citep{zimmerman+23, hiramatsu+23}, the mass-loss rate is found to be of the order of $10^{-2}$\,\mdot,  indicating the forward shock is indeed radiative around day 4. 
This value for the mass-loss rate is  $\ge10^{-3}$\,\mdot\ and
$\ge 5\times 10^{-3}$\,\mdot\ on day 13 and 86 for the forward shock to be radiative.  Our mass-loss rate estimation derived above at the \chandra\ epochs considering adiabatic expression is self-consistent.

Assuming the column density is due to the CSM  provides an independent estimate of the mass loss rate. Following \citet{fransson+96}, this value is $8.7  \times 10^{-5}$ \mdot\ at the two epochs, which is clearly smaller than the one obtained from the X-ray luminosity. The discrepancy is by a factor of 6. Even if  the metallicity is half of solar metallicity, the column density will increase by only  a factor of 2 and  the discrepancy cannot be alleviated. 
This 
%can probably be ascribed to a non-spherically symmetrics CSM.   
may be due to several factors like, non-spherically symmetric CSM, clumpiness, lack of electron-ion equipartition etc \citep{cf17}.

Based on our derived mass-loss rate and forward shock velocity, we can constrain the forward shock radius  at the two \chandra\ epochs to be $0.56 \times 10^{15}$\,cm and $3.72 \times 10^{15}$\,cm,  and the CSM density at these epochs to be $7.84\times 10^{-16}$\,g\,cm$^{-3}$ and 
$1.79 \times 10^{-17}$\,g\,cm$^{-3}$. These values are in excellent agreement with mapping of the CSM structure provided by \citet{zimmerman+23}.

The most likely origin of the 6.4 keV line at the first epoch, which we identified with the Fe K$\alpha$ line, is via the fluorescent reprocessing of  X-rays by the cold CSM  \citep{makishima86}.  In this scenario the equivalent width depends upon the geometry  and the iron abundance \citep{matt+91}, and hence a good indicator of the column density. Under this scenario, a simple linear scaling between the equivalent width and column density is expected \citep{torrejon+10}. The equivalent width in our \chandra\ measurements is 0.8 keV, which implies much larger column density than indicated by the observations.
This could be due to non-spherically symmetric medium as well as the line arising from Fe of different ionization states covering wider region, partially ionized plasma and farther away neutral medium \citep{makishima86}.

An issue to discuss is the following:  the Iron K-alpha line arises from the low ionization of Iron (i.e. FeI to Fe XIX). For this the ionization parameter should be $\le 100$ \citep{fabian+00}. Assuming the CSM is ionized by the X-rays, we can estimate the ionization 
parameter $\zeta = L/n r^2$ using the X-ray luminosity and our derived mass-loss rate. 
The mass-loss rate derived from the X-ray luminosity corresponds to a number density of $\sim 3.7 \times 10^8$ cm$^{-3}$.
The expected ionization parameter is then around 250 at the first \chandra\ epoch. This value means that elements C, N, and O are completely ionized but heavy elements like S and Fe are not. However, at this ionization parameter, iron is likely to be ionized beyond Fe XIX as well, and will show a shift in the centroid energy, which has not been seen in our measurements. This could possibly be due to asymmetric distribution of the CSM, which has also been indicated by the spectropolarimetric data \citep{vasylyev+23} as well as the  lack of narrow blueshifted absorption  profiles in optical spectra \citep{smith+23}. 

The non-detection of Fe K$alpha$ line at the second epoch is consistent with the scenario where line originates from reprocessing of X-ray radiation via cold CSM. At the second epoch, $N_H$ is smaller by a factor of 7, so the strength of the Fe line should be correspondingly smaller. We can make this consistency check. Under the assumption that the equivalent width of the  Fe line did not change at the second epoch,  the calculated  line flux will be $1.13\times^{-14}$\,\flux. 
At the second epoch the column density has decreased by a factor of 7. Thus a factor of 7 decrease in line strength will imply a line  flux of $1.4\times10^{-14}$\,\flux, which is consistent with the calculated line flux assuming constant equivalent width. 

\section{Summary and conclusions}

In this work we present \snixf\ observations with the  \chandra\ ACIS-S at two epochs separated by around 2.5 months.  The X-ray luminosity is  arising from the adiabatic forward shock and its value ($\sim 10^{40}$\,\lum)  falls somewhere in the average luminosity of core-collapse SNe \citep{chandra18}. 
The column density and the 0.3--10\,keV luminosity decline as $t^{-1}$ during the two \chandra\ observations on day 13 and 86, indicating that the SN ejecta is expanding in the standard $r^{-2}$ wind created by a constant mass-loss rate of $(5.6\pm 0.18) \times 10^{-4}$\,\mdot\ during 10--1.5 years before explosion, and is consistent with the \snixf\ progenitor being a   massive red supergiant. 
%\km{What do you mean by `a more massive RSG'? Perhaps rephrase it - simply `a massive RSG' or `a massive-end of an RSG' or something like that?}
The day 13 \chandra\ data also show the presence of the
Fe 6.4 keV line suggesting that the CSM may not have been fully ionized by the X-ray radiation. The line disappears on the second epoch, consistent with the decreased column density. The X-ray data indicate asymmetry in the CSM.

\snixf\ will be monitored for years to come and future studies will throw light on our understanding of massive stars evolving to become CCSNe at their endpoints.

 \section{Acknowledgments}
 We thank the referee for very thoughtful comments.  PC acknowledges support from NASA through Chandra award number DD3-24141X issued by the Chandra X-ray Center. RAC acknowledges support from NSF grant AST-1814910.
K.M. acknowledges support from the Japan Society for the Promotion of Science (JSPS) KAKENHI grant JP18H05223, JP20H00174, and JP20H04737.
 The scientific results reported in this article are based on
 observations made by the Chandra X-ray Observatory, data obtained from the Chandra Data Archive, and software provided by the Chandra X-ray Center.
 The National Radio Astronomy Observatory is a facility of the National Science Foundation operated under cooperative agreement by Associated Universities, Inc.
 This research has made use of NASA's Astrophysics Data System Bibliographic Services. This research has made use of data and/or software provided by the High Energy Astrophysics Science Archive Research Center (HEASARC), which is a service of the Astrophysics Science Division at NASA/GSFC. We acknowledge the use of public data from the Swift data archive.\\
 This paper employs a list of \chandra\, datasets, obtained by the \chandra\, X-ray Observatory, contained in~\dataset[DOI: https://doi.org/10.25574/cdc.196]{https://doi.org/10.25574/cdc.196}.
 
\bibliography{ms}{}
\bibliographystyle{aasjournal}

\end{document}